\documentclass{PoS}

\usepackage{amsmath}
\usepackage{bm}
\usepackage{slashed}

\allowdisplaybreaks

\title{The Symmetries of the Three Heavy Quark Bound State and the Singlet Static Energy}

\ShortTitle{The Symmetries of the Three Heavy Quark Bound State and the Singlet Static Energy}

\author{\speaker{Felix Karbstein}\\
Helmholtz-Institut Jena, Fr\"obelstieg 3, D-07743 Jena, Germany \&\\
Theoretisch-Physikalisches Institut, Friedrich-Schiller-Universit\"at Jena, Max-Wien-Platz 1,\\ D-07743 Jena, Germany\\
E-mail: \email{f.karbstein@gsi.de}}

\abstract{
Potential non-relativistic QCD provides a convenient framework to study the three heavy-quark system.
We have studied its symmetries under exchange of the heavy quarks and determined the leading ultrasoft contributions to the singlet static potential and to the singlet static energy.

Specializing to an equilateral geometry, we have solved the renormalization group equations in order to resum the leading ultrasoft logarithms for the singlet static potential, which is thus known at NNLL accuracy. 
}

\FullConference{Xth Quark Confinement and the Hadron Spectrum\\
		 8-12 October 2012\\
		 TUM Campus Garching, Munich, Germany}

\begin{document}

\section{Introduction}

We have studied the symmetries of the three heavy-quark system under exchange 
of the heavy-quark fields and their implications for the various matching coefficients, i.e. the potentials, of potential non-relativistic QCD (pNRQCD) for the three heavy-quark system.
Moreover, we have calculated the ultrasoft corrections of order $\alpha_{\rm s}^4\ln\alpha_{\rm s}$ 
to the singlet static energy and of order $\alpha_{\rm s}^4\ln\mu$ to the singlet static potential of a three heavy-quark bound state.
Whereas this has been achieved for the case of $Q\bar Q$ systems more than ten years ago~\cite{Brambilla:1999qa}, 
the result for three heavy-quark systems are new \cite{wir}.

\section{Three heavy-quark composite fields}\label{sec:QQQ}

Quarks transform under the fundamental representation, $3$, of the (color) gauge group ${\rm SU}(3)_c$.
According to
\begin{equation}
 3\otimes3\otimes3=1\oplus8\oplus8\oplus10\,, \label{prodinirreps}
\end{equation}
the three heavy-quark product state, $Q_i({\bf x}_1,t)Q_j({\bf x}_2,t)Q_k({\bf x}_3,t)$ ($i,j,k=1,2,3$ denote color indices), can thus be decomposed into a singlet, two different octets and a decuplet
with respect to ${\rm SU}(3)_c$ gauge transformations at a common point $\bf R$.
Employing equal-time straight Wilson strings,
\begin{align}
\phi({\bf y},{\bf x},t)=
{\cal P}\exp\left\{ig\int_0^1{\rm d}s\ ({\bf y}-{\bf x})\cdot{\bf A}({\bf x}+({\bf y}-{\bf x})s,t)\right\},
\end{align}
where ${\bf A}={\bf A}^a\lambda^a/2$ is the color gauge field, $\lambda^a$ are the Gell-Mann matrices, and ${\cal P}$ denotes path ordering of the color matrices,
we write
\begin{multline}
 Q_{i}({\bf x}_1,t)Q_{j}({\bf x}_2,t)Q_{k}({\bf x}_3,t) 
=\phi_{ii'}({\bf x}_1,{\bf R},t)\phi_{jj'}({\bf x}_2,{\bf R},t)\phi_{kk'}({\bf x}_3,{\bf R},t) \\
\times\biggr\{S({\bf x}_1,{\bf x}_2,{\bf x}_3,t){\underline {\bf S}}_{i'j'k'}
+\sum_{a=1}^8O^{Aa}({\bf x}_1,{\bf x}_2,{\bf x}_3,t){\underline {\bf O}}^{Aa}_{i'j'k'}\\+\sum_{a=1}^8O^{Sa}({\bf x}_1,{\bf x}_2,{\bf x}_3,t){\underline {\bf O}}^{Sa}_{i'j'k'}
+\sum_{\delta=1}^{10}\Delta^{\delta}({\bf x}_1,{\bf x}_2,{\bf x}_3,t){\underline {\pmb \Delta}}^{\delta}_{i'j'k'}\biggl\},
\label{eq:dec}
\end{multline}
where ${\underline {\bf S}}_{ijk}$, ${\underline {\bf O}}^{Aa}_{ijk}$,  ${\underline {\bf O}}^{Sa}_{ijk}$ and 
${\underline {\pmb \Delta}}^{\delta}_{ijk}$ are orthogonal and normalized  color tensors that satisfy the relations 
\begin{align}
&{\underline {\bf S}}_{ijk}{\underline {\bf S}}_{ijk}=1\,, \quad {\underline {\bf O}}^{Aa*}_{ijk}{\underline {\bf O}}^{Ab}_{ijk}
=\delta^{ab}\,, \quad {\underline {\bf O}}^{Sa*}_{ijk}{\underline {\bf O}}^{Sb}_{ijk}
=\delta^{ab}\,, \quad {\underline {\pmb \Delta}}^{\delta}_{ijk}{\underline {\pmb \Delta}}^{\delta'}_{ijk}=\delta^{\delta\delta'}\,, 
\nonumber\\
&{\underline {\bf S}}_{ijk}{\underline {\bf O}}^{Aa}_{ijk}={\underline {\bf S}}_{ijk}{\underline {\bf O}}^{Sa}_{ijk}
={\underline {\bf S}}_{ijk}{\underline {\pmb \Delta}}^{\delta}_{ijk}
={\underline {\bf O}}^{Aa*}_{ijk}{\underline {\bf O}}^{Sb}_{ijk}
={\underline {\bf O}}^{Aa*}_{ijk}{\underline {\pmb \Delta}}^{\delta}_{ijk}
={\underline {\bf O}}^{Sa*}_{ijk}{\underline {\pmb \Delta}}^{\delta}_{ijk}=0\,, \label{OrthoN}
\end{align}
with $a,b\in \{1, \ldots, 8\}$, and $\delta,\delta'\in \{1, \ldots, 10\}$~\cite{Brambilla:2005yk}.
For simplicity, we have omitted an explicit reference to ${\bf R}$ in the argument of the singlet, $S$, two octets, $O^A$ and $O^S$, and decuplet, $\Delta$, fields.
Besides the time coordinate $t$, we rather list the position coordinates 
(${\bf x}_1,{\bf x}_2,{\bf x}_3$) of the heavy-quark fields in the order (from left to right) 
of their appearance on the right-hand side of Eq.~(\ref{eq:dec}). 
The same convention is used for the color indices ($i,j,k$).

The heavy fields fulfill equal-time anticommutation relations, 
\begin{align}
\{Q_{i}({\bf x},t),Q_{j}({\bf y},t)\}=0. \label{vert}
\end{align}
Thus, it is natural to ask for the implications of Eq.~\eqref{vert} for the singlet, octet and decuplet fields in Eq.~\eqref{eq:dec}.
In order to do this, we stick to a specific (matrix) representation of the color tensors, namely \cite{Brambilla:2005yk}
\begin{equation}
{\underline {\bf S}}_{ijk}=\frac{1}{\sqrt{6}}\epsilon_{ijk}\,,
\quad {\underline {\bf O}}^{Aa}_{ijk}=\frac{1}{2}\epsilon_{ijl}\lambda^a_{kl}\,, \quad {\underline {\bf O}}^{Sa}_{ijk}=\frac{1}{2\sqrt{3}}\left(\epsilon_{jkl}\lambda^a_{il}+\epsilon_{ikl}\lambda^a_{jl}\right)\,,
\end{equation}
and
\begin{align}
{\underline {\pmb \Delta}}^{1}_{111}&={\underline {\pmb \Delta}}^{4}_{222}={\underline {\pmb \Delta}}^{10}_{333}=1\,,
\quad\quad {\underline {\pmb \Delta}}^{6}_{\{123\}}=\frac{1}{\sqrt{6}}\,, \nonumber\\
{\underline {\pmb \Delta}}^{2}_{\{112\}}&={\underline {\pmb \Delta}}^{3}_{\{122\}}={\underline {\pmb \Delta}}^{5}_{\{113\}}
={\underline {\pmb \Delta}}^{7}_{\{223\}}={\underline {\pmb \Delta}}^{8}_{\{133\}}={\underline {\pmb \Delta}}^{9}_{\{233\}}
=\frac{1}{\sqrt{3}}\,, \label{Delta}
\end{align}
where the symbol $\{ijk\}$ denotes all permutations of the indices $ijk$; all components not listed explicitly in Eq.~(\ref{Delta}) are zero.
Moreover, note that as all six possible orderings of the heavy-quark fields can be obtained by applying the following two independent operations on the product state $Q_i({\bf x}_1,t)Q_j({\bf x}_2,t)Q_k({\bf x}_3,t)$,
\begin{equation}
 (i):\quad Q_i({\bf x}_1,t)\leftrightarrow Q_j({\bf x}_2,t)\,, \quad (ii):\quad Q_i({\bf x}_1,t)\leftrightarrow Q_k({\bf x}_3,t)\,,
\end{equation}
it suffices to determine the transformation behavior of $S$, $O^{Aa}$, $O^{Sa}$ and $\Delta$ under $(i)$ and $(ii)$.
We find
\begin{align}
(i):\quad
\begin{cases}
S({\bf x}_1,{\bf x}_2,{\bf x}_3,t)             \hspace{-3mm} &= \phantom{-}S({\bf x}_2,{\bf x}_1,{\bf x}_3,t) \\
\Delta^{\delta}({\bf x}_1,{\bf x}_2,{\bf x}_3,t) \hspace{-3mm} &=  -\Delta^{\delta}({\bf x}_2,{\bf x}_1,{\bf x}_3,t) \\
O^{Aa}({\bf x}_1,{\bf x}_2,{\bf x}_3,t)         \hspace{-3mm} &=  \phantom{-}O^{Aa}({\bf x}_2,{\bf x}_1,{\bf x}_3,t) \\
O^{Sa}({\bf x}_1,{\bf x}_2,{\bf x}_3,t)         \hspace{-3mm} &=  -O^{Sa}({\bf x}_2,{\bf x}_1,{\bf x}_3,t) 
\label{block1}
\end{cases}
\,,
\end{align}
and 
\begin{align}
(ii):\quad
\begin{cases}
S({\bf x}_1,{\bf x}_2,{\bf x}_3,t)             \hspace{-3mm} &=  \phantom{-}S({\bf x}_3,{\bf x}_2,{\bf x}_1,t)\\
\Delta^{\delta}({\bf x}_1,{\bf x}_2,{\bf x}_3,t) \hspace{-3mm} &= -\Delta^{\delta}({\bf x}_3,{\bf x}_2,{\bf x}_1,t)\\
O^{Aa}({\bf x}_1,{\bf x}_2,{\bf x}_3,t)         \hspace{-3mm} &= 
-\cos(\tfrac{\pi}{3})O^{Aa}({\bf x}_3,{\bf x}_2,{\bf x}_1,t) + \sin(\tfrac{\pi}{3})O^{Sa}({\bf x}_3,{\bf x}_2,{\bf x}_1,t)\\
O^{Sa}({\bf x}_1,{\bf x}_2,{\bf x}_3,t)         \hspace{-3mm} &= 
\phantom{-}\sin(\tfrac{\pi}{3})O^{Aa}({\bf x}_3,{\bf x}_2,{\bf x}_1,t) + \cos(\tfrac{\pi}{3})O^{Sa}({\bf x}_3,{\bf x}_2,{\bf x}_1,t)
\label{block2}
\end{cases}
\,. 
\end{align}

\section{The pNRQCD Lagrangian for the three heavy-quark system}

As pNRQCD for the three heavy-quark system is usually formulated in terms of singlet, octet and decuplet fields, the symmetry relations~\eqref{block1} and \eqref{block2}
have some immediate consequences for the form of the pNRQCD Lagrangian, which is organized as a double expansion in $1/m$ and in the relative coordinates ${\bf r}_i$ ($i=1,2,3$).
To zeroth order in both expansions, it reads \cite{Brambilla:2005yk}
\begin{eqnarray}
{\cal L}_{\rm pNRQCD}^{(0,0)}&=&\int{\rm d}^3\!\rho\,{\rm d}^3\!\lambda\;\Bigl\{
S^{\dag}\left[i\partial_0-V^s\right]S+\Delta^{\dag}\left[iD_0-V^{d}\right]\Delta+O^{A\dag}\left[iD_0-V^o_A\right]O^A 
\nonumber\\
&&\hspace*{1.7cm}+O^{S\dag}\left[iD_0-V^o_S\right]O^S +O^{A\dag}\left[-V_{AS}^o\right]O^S+O^{S\dag}\left[-V_{AS}^o\right]O^A\Bigr\} 
\nonumber\\
&&+\sum_{l}\bar{q}^{\,l}i\slashed{D}q^l-\frac{1}{4}F^a_{\mu\nu}F^{a\mu\nu}\,.
\label{LpNRQCD1}
\end{eqnarray}
Equation~\eqref{LpNRQCD1} describes at zeroth order in the multipole expansion the propagation of light quarks and ultrasoft gluons as well as the temporal evolution of static quarks.
Higher order terms, not displayed explicitly here, account for corrections due to finite heavy-quark masses and interactions between heavy-quarks and ultrasoft gluons.
As there is no preferred ordering of the heavy-quarks, Eq.~\eqref{LpNRQCD1} has to be invariant under different orderings of the heavy-quark fields.

The potentials $V$ can be expressed in terms of the relative vectors
\begin{equation}
{\bf r}_1={\bf x}_1-{\bf x}_2\,, \qquad
{\bf r}_2={\bf x}_1-{\bf x}_3\,, \qquad
{\bf r}_3={\bf x}_2-{\bf x}_3\,.
\label{rx123}
\end{equation}
For the ordering of the heavy-quarks as in Eq.~\eqref{eq:dec}, i.e. $S\equiv S({\bf x}_1,{\bf x}_2,{\bf x}_3,t)$, $O^A\equiv O^A({\bf x}_1,{\bf x}_2,{\bf x}_3,t)$, $O^S\equiv O^S({\bf x}_1,{\bf x}_2,{\bf x}_3,t)$   
and $\Delta\equiv \Delta({\bf x}_1,{\bf x}_2,{\bf x}_3,t)$, they are defined as $V\equiv V({\bf r}_1,{\bf r}_2,{\bf r}_3)$.
Given the symmetry relations for the singlet, octet and decuplet fields, Eqs.~\eqref{block1} and \eqref{block2}, it is straightforward to also derive corresponding symmetry relations for the potentials in Eq.~\eqref{LpNRQCD1}.
The singlet and decuplet potentials remain invariant under $(i)$ and $(ii)$, whereas the octet potentials transform as
\begin{align}
(i):\quad
\begin{cases}
V^o_A({\bf r}_1,{\bf r}_2,{\bf r}_3)     \hspace{-3mm} &=  \phantom{-} V^o_A(-{\bf r}_1,{\bf r}_3,{\bf r}_2) \\
V^o_S({\bf r}_1,{\bf r}_2,{\bf r}_3)     \hspace{-3mm} &=  \phantom{-} V^o_S(-{\bf r}_1,{\bf r}_3,{\bf r}_2) \\
V_{AS}^o({\bf r}_1,{\bf r}_2,{\bf r}_3)   \hspace{-3mm} &=  - V_{AS}^o(-{\bf r}_1,{\bf r}_3,{\bf r}_2)
\label{Vrel1}
\end{cases}\,, 
\end{align}
and 
\begin{align}
(ii):\quad
\begin{pmatrix}
 V^o_A({\bf r}_1,{\bf r}_2,{\bf r}_3)  \\
 V^o_S({\bf r}_1,{\bf r}_2,{\bf r}_3)  \\
 V_{AS}^o({\bf r}_1,{\bf r}_2,{\bf r}_3)
\end{pmatrix}
=
\begin{pmatrix}
 \cos^2(\tfrac{\pi}{3}) & \sin^2(\tfrac{\pi}{3}) & -\sin(\tfrac{2\pi}{3}) \\
 \sin^2(\tfrac{\pi}{3}) & \cos^2(\tfrac{\pi}{3}) & +\sin(\tfrac{2\pi}{3}) \\
 \frac{1}{2}\sin(\tfrac{2\pi}{3}) & -\frac{1}{2}\sin(\tfrac{2\pi}{3}) & -\cos(\tfrac{2\pi}{3})
\end{pmatrix}
\begin{pmatrix}
 V^o_A(-{\bf r}_3,-{\bf r}_2,-{\bf r}_1)  \\
 V^o_S(-{\bf r}_3,-{\bf r}_2,-{\bf r}_1)  \\
 V_{AS}^o(-{\bf r}_3,-{\bf r}_2,-{\bf r}_1)
\end{pmatrix}.
\label{Vrel2}
\end{align}
Similar relations hold, e.g. for the interaction vertices in the pNRQCD Lagrangian appearing at higher order in the multipole expansion in the relative vectors ${\bf r}_i$ ($i=1,2,3$).

The symmetry properties of the three heavy-quark system under exchange of the heavy-quark fields thus have a deep impact on the structure of the pNRQCD Lagrangian.
They in particular induce relations between different matching coefficients in the effective theory and thereby constrain their form.

\section{The singlet static energy up to order $\alpha_s^4\ln\alpha_s$}

Besides studying the symmetries of the three heavy-quark system, we have used the effective field theory framework of pNRQCD to determine the leading ultrasoft contribution to the singlet static energy,
which is of $\alpha_s^4\ln\alpha_s$, and to the singlet static potential, which is of order $\alpha_s^4\ln\mu$. Here, the symmetry relations~\eqref{Vrel1} and \eqref{Vrel2} have served as an important check of the obtained result.

Adding the newly determined leading ultrasoft corrections to the singlet static potential $V^s$, known analytically at next-to-next-to-leading order (NNLO) \cite{Brambilla:2009cd},
the singlet static energy is now known up to order $\alpha_{\rm s}^4\ln\alpha_{\rm s}$ and reads
\begin{eqnarray}
E^s({\bf r}_1,{\bf r}_2,{\bf r}_3) &=& V^s_{\rm NNLO}({\bf r}_1,{\bf r}_2,{\bf r}_3)
\nonumber\\
- \frac{\alpha_{\rm s}^4}{3\pi}\ln\alpha_{\rm s} && \hspace{-6mm}
\left[
\left({\bf r}_1^2+\frac{({\bf r}_2+{\bf r}_3)^2}{3}\right)
\left(\frac{1}{|{\bf r}_1|^2} + \frac{1}{|{\bf r}_2|^2} + \frac{1}{|{\bf r}_3|^2} 
-\frac{1}{4}\frac{|{\bf r}_1| + |{\bf r}_2| + |{\bf r}_3|}{|{\bf r}_1||{\bf r}_2||{\bf r}_3|}\right)
\right.
\nonumber\\
&&
\hspace{3.4cm}
\times \left(\frac{1}{|{\bf r}_1|} + \frac{1}{|{\bf r}_2|} + \frac{1}{|{\bf r}_3|} \right)
\nonumber\\
&&
\hspace{-6mm}
+
\left({\bf r}_1^2-\frac{({\bf r}_2+{\bf r}_3)^2}{3}\right)
\left(\frac{1}{|{\bf r}_1|^2} + \frac{1}{|{\bf r}_2|^2} + \frac{1}{|{\bf r}_3|^2} 
+\frac{5}{4}\frac{|{\bf r}_1| + |{\bf r}_2| + |{\bf r}_3|}{|{\bf r}_1||{\bf r}_2||{\bf r}_3|}\right)
\nonumber\\
&&
\hspace{3.4cm}
\times \left(\frac{1}{|{\bf r}_1|} - \frac{1}{2|{\bf r}_2|} - \frac{1}{2|{\bf r}_3|} \right)
\nonumber\\
&&
\hspace{-6mm}
+
{\bf r}_1\cdot({\bf r}_2+{\bf r}_3)
\left(\frac{1}{|{\bf r}_1|^2} + \frac{1}{|{\bf r}_2|^2} + \frac{1}{|{\bf r}_3|^2} 
+\frac{5}{4}\frac{|{\bf r}_1| + |{\bf r}_2| + |{\bf r}_3|}{|{\bf r}_1||{\bf r}_2||{\bf r}_3|}\right)
\nonumber\\
&&
\hspace{3.4cm}
\left.
\times \left(\frac{1}{|{\bf r}_2|} - \frac{1}{|{\bf r}_3|} \right)
\right]
\,.
\label{eq:E0full}
\end{eqnarray}

In contrast to the static energy, the singlet static potential explicitly depends on the factorization scale $\mu$ separating soft from ultrasoft contributions. It is now known up to order $\alpha_s^4\ln\mu$, where the quantity $\ln\mu$ is in general referred to as an ultrasoft logarithm.
In a minimal subtraction scheme, it is given by
\begin{eqnarray}
V^s({\bf r}_1,{\bf r}_2,{\bf r}_3;\mu) &=& V^s_{\rm NNLO}({\bf r}_1,{\bf r}_2,{\bf r}_3)
\nonumber\\
- \frac{\alpha_{\rm s}^4}{3\pi}\ln\mu && \hspace{-6mm}
\left[
\left({\bf r}_1^2+\frac{({\bf r}_2+{\bf r}_3)^2}{3}\right)
\left(\frac{1}{|{\bf r}_1|^2} + \frac{1}{|{\bf r}_2|^2} + \frac{1}{|{\bf r}_3|^2} 
-\frac{1}{4}\frac{|{\bf r}_1| + |{\bf r}_2| + |{\bf r}_3|}{|{\bf r}_1||{\bf r}_2||{\bf r}_3|}\right)
\right.
\nonumber\\
&&
\hspace{3.4cm}
\times \left(\frac{1}{|{\bf r}_1|} + \frac{1}{|{\bf r}_2|} + \frac{1}{|{\bf r}_3|} \right)
\nonumber\\
&&
\hspace{-6mm}
+
\left({\bf r}_1^2-\frac{({\bf r}_2+{\bf r}_3)^2}{3}\right)
\left(\frac{1}{|{\bf r}_1|^2} + \frac{1}{|{\bf r}_2|^2} + \frac{1}{|{\bf r}_3|^2} 
+\frac{5}{4}\frac{|{\bf r}_1| + |{\bf r}_2| + |{\bf r}_3|}{|{\bf r}_1||{\bf r}_2||{\bf r}_3|}\right)
\nonumber\\
&&
\hspace{3.4cm}
\times \left(\frac{1}{|{\bf r}_1|} - \frac{1}{2|{\bf r}_2|} - \frac{1}{2|{\bf r}_3|} \right)
\nonumber\\
&&
\hspace{-6mm}
+
{\bf r}_1\cdot({\bf r}_2+{\bf r}_3)
\left(\frac{1}{|{\bf r}_1|^2} + \frac{1}{|{\bf r}_2|^2} + \frac{1}{|{\bf r}_3|^2} 
+\frac{5}{4}\frac{|{\bf r}_1| + |{\bf r}_2| + |{\bf r}_3|}{|{\bf r}_1||{\bf r}_2||{\bf r}_3|}\right)
\nonumber\\
&&
\hspace{3.4cm}
\left.
\times \left(\frac{1}{|{\bf r}_2|} - \frac{1}{|{\bf r}_3|} \right)
\right]
\,.
\label{Vs3loop}
\end{eqnarray}
Specializing to an equilateral geometry, characterized by the single length scale $r=|{\bf r}_1|=|{\bf r}_2|=|{\bf r}_3|$, we have moreover managed to resum the leading ultrasoft logarithms that start appearing in the static potential at NNNLO to all orders by solving the corresponding renormalization group equations.
For the singlet static potential this results in
\begin{equation}
 V^s(r;\mu) = V^s_{\rm NNLO}(r)-9\frac{\alpha_{\rm s}^3(1/r)}{\beta_0r} \ln\frac{\alpha_{\rm s}(1/r)}{\alpha_{\rm s}(\mu)}\,,
\label{vsr}
\end{equation}
where $\beta_0 = 11 -2/3n_l$, with $n_l$ the number of light-quark flavors.
Equation~\eqref{vsr} provides the complete expression of the singlet static potential at next-to-next-to-leading-logarithmic (NNLL) accuracy in an equilateral geometry. Corresponding results for the octet and decuplet potentials can be found in \cite{wir}.

\subsection*{Acknowledgments}

The author gratefully acknowledges collaboration with N.~Brambilla and A.~Vairo on the presented topics, as well as financial support from the FAZIT foundation and from the DFG cluster of excellence ``Origin and structure of the universe'' (www.universe-cluster.de).

\end{document}